\documentclass[twocolumn,showpacs,floatfix,prl,superscriptaddress]{revtex4}

\usepackage{float}
\usepackage{graphicx}
\usepackage{color}
\usepackage{hyperref}

\newcommand{\beq}{\begin{eqnarray}}
\newcommand{\eeq}{\end{eqnarray}}
\newcommand{\beqq}{\begin{eqnarray*}}
\newcommand{\eeqq}{\end{eqnarray*}}

\begin{document}

\title{Tensor Network Implementation of Bulk Entanglement Spectrum}

\author{Timothy H. Hsieh}
\affiliation{Department of Physics, Massachusetts Institute of Technology, Cambridge, MA 02139}
\author{Liang Fu}
\affiliation{Department of Physics, Massachusetts Institute of Technology, Cambridge, MA 02139}
\author{Xiao-Liang Qi}
\affiliation{Department of Physics, Stanford University, Stanford, California 94305, USA.}

\begin{abstract}
Many topologically nontrivial states of matter possess gapless degrees of freedom on the boundary, and when these boundary states delocalize into the bulk, a phase transition occurs and the system becomes topologically trivial.  We show that tensor networks provide a natural framework for analyzing such topological phase transitions in terms of the boundary degrees of freedom which mediate it.  To do so, we make use of a correspondence between a topologically nontrivial ground state and its phase transition to a trivial phase established in \cite{BES}.  This involved computing the bulk entanglement spectrum (BES) of the ground state, upon tracing out an extensive subsystem.  This work implements BES via tensor network representations of ground states.  In this framework, the universality class of the quantum critical entanglement Hamiltonian in $d$ spatial dimensions is either derived analytically or mapped to a classical statistical model in $d+1$ dimensions, which can be studied using Monte Carlo or tensor renormalization group methods.  As an example, we analytically derive the universality classes of topological phase transitions from the spin-1 chain Haldane phase and demonstrate that the AKLT wavefunction (and its generalizations) remarkably contains critical six-vertex (and in general eight-vertex) models within it.
\end{abstract}


\maketitle

Despite lacking a local order parameter, topological states contain a wealth of subtly encoded information \cite{
tknn, hasankane, qizhang, moore, niuwen, wen, kanemele, nayak}, including in some cases topological invariants such as Chern number, in other cases ground state degeneracy on higher genus manifolds, for example.  Given the stark contrast between topological states and classically ordered states, it is natural that purely quantum notions are often necessary for analyzing topological states.  In particular, measures of entanglement, which has no classical analog, have proven to be extremely useful.  For example, the entanglement entropy between a subsystem and its complement has been used \cite{levinwen, kitaev, hamma1, hamma2, zhang} to detect topological order in a ground state.  Moreover, the full spectrum of the reduced density matrix, called the entanglement spectrum \cite{lihaldane, locality, bernevig, qi, swingle, pollman, fid, yao, dubail}, has allowed one to simulate the edge excitations of a topological ground state.  In other words, tracing out a subsystem from a ground state achieves a similar effect as introducing excitations localized at the boundary of the subsystem.

With this paradigm in mind, two of us have introduced a new technique called bulk entanglement spectrum (BES) to study the {\it bulk} of a system \cite{BES}.  Specifically, it was found that a topological state contains information about its phase transition to a trivial phase.  How is it possible that a single wavefunction can give birth to quantum criticality associated with a topological phase transition?  By using a special partition of the topological state that is extensive in all directions and possesses symmetry between the remaining and traced out subsystems, the resulting bulk entanglement Hamiltonian (see eq.\ref{EH}) was argued to be either critical or possess ground state degeneracy.  In the former case, the critical bulk entanglement Hamiltonian sits right at a phase transition between the original topologically nontrivial phase and a trivial phase.  The essence of this argument is the discrete nature of topological order and extreme limits of the partition.  When nearly nothing is traced out, the remaining subsystem is in the same topological phase as the original nontrivial wave function.  When nearly everything is traced out, the remaining subsystem consists of decoupled small ``islands" and thus is topologically trivial.  Tuning the geometry of the partition thus induces a phase transition in the entanglement Hamiltonian.  If there is a single phase transition, then it must occur at the intermediate, symmetric partition defined above.  This protocol, which is derived from a single wavefunction, differs markedly from the usual realization of a quantum phase transition by tuning parameters in a Hamiltonian.  As a proof of concept, BES has been rigorously shown to work for integer quantum Hall states.  The BES produces the massless Dirac spectrum expected at the transition between states with Chern numbers one and zero.


In this paper, we study the quantum phase transition in BES for generic matrix product states (MPS)\cite{mps, tns}, which are efficient representations of generic non-critical states in one dimension. We analytically implement the BES technique and obtain a more explicit understanding of the entanglement Hamiltonian and topological phase transition.
The partition function of this bulk entanglement Hamiltonian serves as the centerpiece of this work, with three primary uses.  First, it allows us to identify the critical theory of the entanglement Hamiltonian.  Second, it maps the quantum critical $d$-dimensional system to a classical $d+1$-dimensional system, enabling Monte Carlo numerics or tensor renormalization group methods \cite{trg} to tackle quantum critical problems.  Finally, it provides a dynamic picture of a topological phase transition and highlights the importance of edge states in mediating such a phase transition.  We will explicitly show that the virtual degree of freedom in the MPS gains a life of its own in the partition function, and it is precisely the interactions of these virtual elements which constitute the topological phase transition.  While we focus on 1d systems in this work, many of our techniques generalize to tensor network wavefunctions in any dimension.



We begin by briefly reviewing BES and tensor networks before using them together.  Consider a ground state $|\Psi \rangle$ defined on a Hilbert space $S$ and partition $S$ into two complementary parts A and B.  After tracing out part B, the description of $|\Psi \rangle$ on A is given by a thermal density matrix
\beq
\rho_A = {\textrm Tr}_B |\Psi\rangle\langle\Psi| \equiv  e^{-H_A }  \label{EH} 
\eeq
corresponding to an entanglement Hamiltonian $H_A$.  The entanglement spectrum is the set of eigenvalues of $H_A$, and in the following we will be interested in the ground state of $H_A$ and its topological nature.

When an {\it extensive partition} is used, i.e. when A and B are extensive with system size in all directions, one attains a {\it bulk} entanglement Hamiltonian: $H_A$ has support on an extensive subsystem $A$.  It was argued in \cite{BES} that when the ground state $|\Psi\rangle$ is an irreducible topological state \cite{irreducible} and when A and B are related by a symmetry, $H_A$ either 1) has ground state degeneracy or 2) is gapless and characterizes a topological phase transition from nontrivial to trivial.  As outlined earlier, this claim can be motivated by considering extreme examples of extensive partitions.  The critical point occurs at the symmetric partition in which A and B are related by a symmetry such as translation.


\begin{figure}
\includegraphics[width=3in]{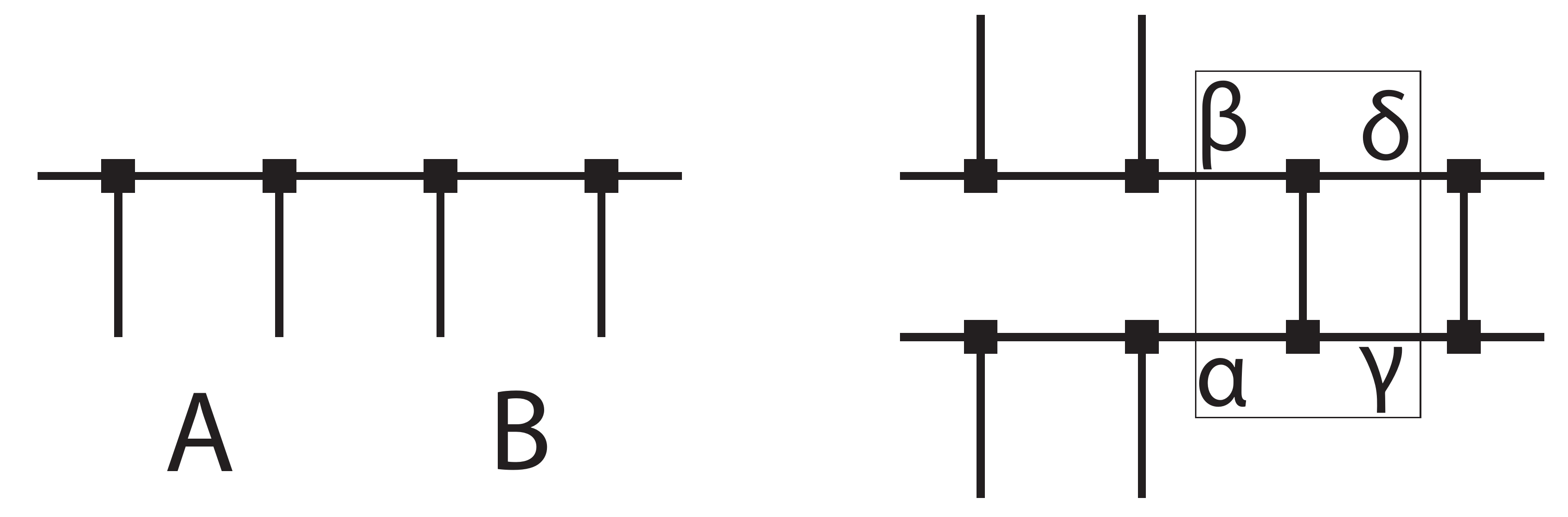}
\caption{(a) A segment of a matrix product state, partitioned into two spatial subspaces A and B. The open-ended vertical links represent physical degrees of freedom and the horizontal links represent `virtual' degrees of freedom which are summed over as in equation 2. (b) The reduced density matrix obtained by tracing out B.  Boxed is the MPS transfer matrix.}
\end{figure}

Tracing out degrees of freedom has a convenient pictorial representation in the framework of tensor networks.  Consider a matrix product state (MPS) given by a tensor $M^{\sigma}_{\{\alpha\}}$ with a physical index $\sigma$ and virtual indices $\{\alpha\}$ emanating from the physical sites (see Fig. 1a).  The virtual indices are contracted, leaving a wavefunction $|\psi\rangle$ defined by
\beq
\langle \sigma_1...\sigma_N|\psi\rangle \equiv \sum_{\text{virtual indices}} M^{\sigma_1} ... M^{\sigma_N}
\eeq

The reduced density matrix $\rho_A$ obtained from tracing out a part $B$ is
\beq
\rho_A = \sum_{\sigma_B} \langle\sigma_B | \psi\rangle\langle \psi|\sigma_B\rangle.
\eeq

Graphically, $\langle \psi|$ is simply represented by reflecting the MPS $|\psi\rangle$ and complex conjugating the tensors.  Then, the $\sigma_B$ indices are contracted to yield $\rho_A$ (Fig. 1b).  In this pictorial language, the topological phase transition realized by tuning the extensive partition is shown in Fig.2, in which the density matrix interpolates between the nontrivial projector onto the topological ground state to a trivial product of density operators.  The phase transition occurs at a partition in which $A$ and $B$ are related by some symmetry.

\begin{figure}
\includegraphics[width=2.5in]{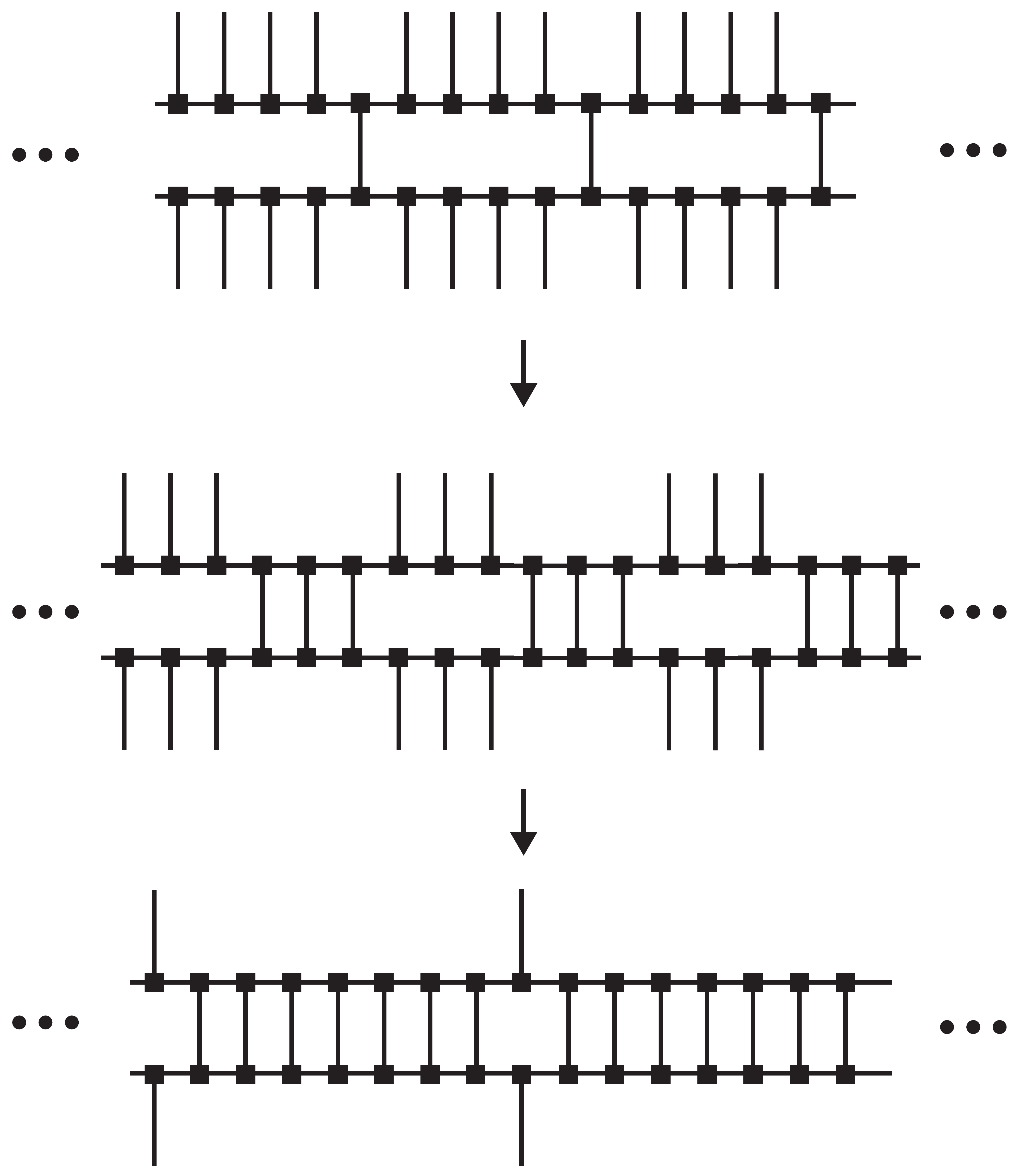}
\caption{The tensor network representation of the BES procedure. The reduced density matrices $\rho_A$ corresponding to different extensive partitions are shown.  Tuning the extensive partition of a topologically nontrivial ground state $\Psi$ realizes a topological phase transition, occurring at the partition where A and B are symmetric (middle).  If too little is traced out (top), $\rho_A\approx |\Psi\rangle\langle \Psi|$ is nontrivial.  If too much is traced out (bottom), $\rho_A \approx \otimes_{i\in A} \rho_i$ is trivial.}
\end{figure}

It is now extremely useful to construct the partition function of the entanglement Hamiltonian $Z=tr(e^{-n H_A}) = tr(\rho_A^n)$, where $n=1/T$ is the inverse `temperature'.  We will eventually take the limit $T\rightarrow 0$ ($n\rightarrow \infty$) to probe the universality class of $H_A$.  Graphically, one simply stacks $n$ copies of $\rho_A$ and then contracts all physical indices, including those at the top of the $n$th copy with those at the bottom of the first copy.  Because all indices are contracted, we now have the freedom to reinterpret the partition function as one involving the virtual degrees of freedom, thus providing a new perspective on the topological phase transition being studied.  In some cases, we can rewrite the partition function as $Z=tr(e^{-\beta \tilde{H}})$, where $\tilde{H}$ now acts on the virtual indices as opposed to the physical indices.  Hence, we call $\tilde{H}$ the effective entanglement Hamiltonian.  Equally importantly, the partition function can be understood as that of a classical model (in one higher dimension) if the Boltzmann weights of all configurations are nonnegative.  The above holds for general extensive partitions, yet it is particularly interesting at the symmetric partition, where the partition function of the critical quantum model can be studied using Monte Carlo or tensor renormalization group methods to attain approximate critical exponents.


\begin{figure}
\includegraphics[width=3in]{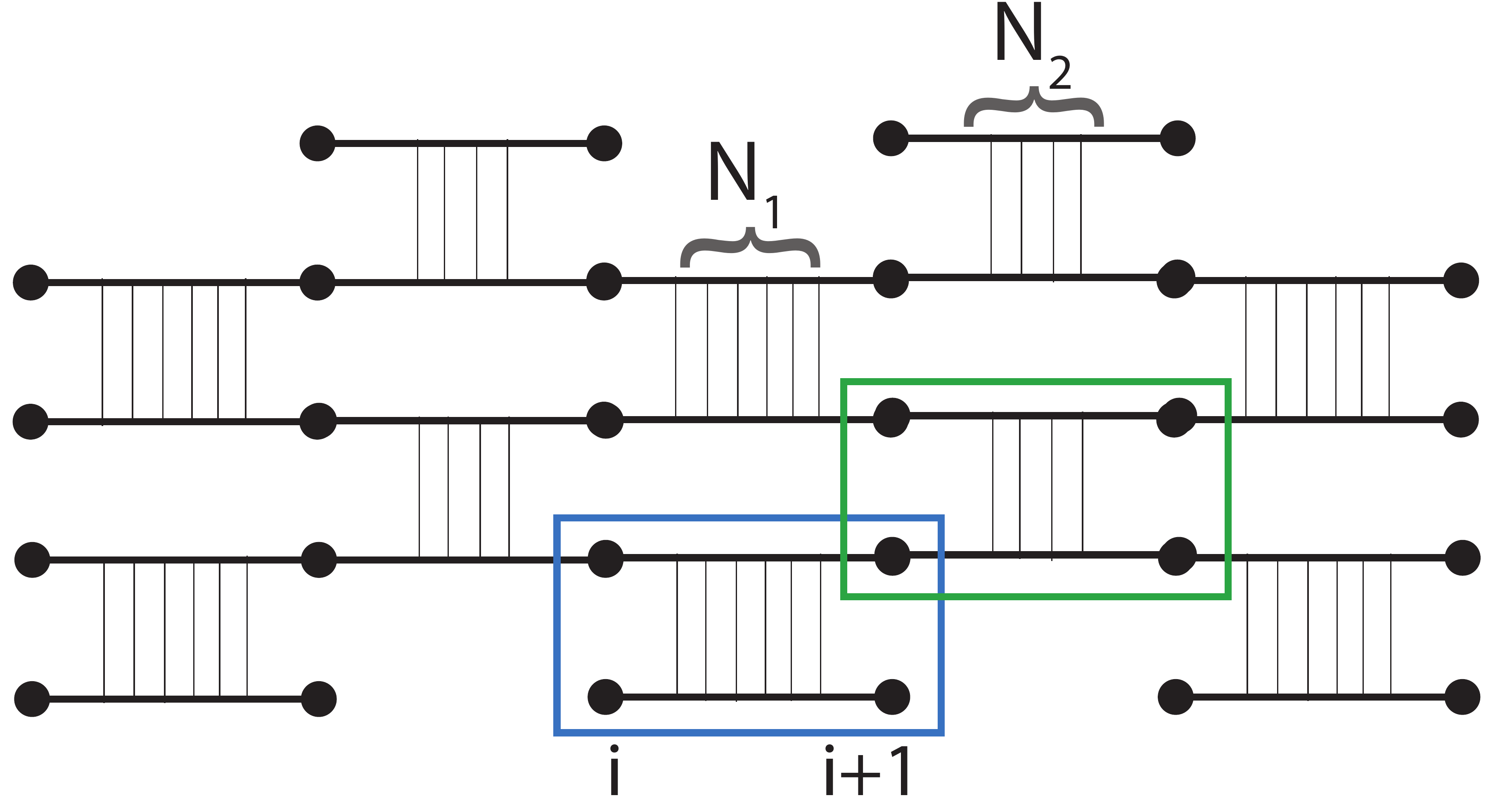}
\caption{The partition function corresponding to the entanglement Hamiltonian from an extensive partition, tracing out blocks of $N_1$ sites and leaving $N_2$ sites in between.  Boxed in blue and green are the building blocks of the partition function, namely the transfer matrices propagating two virtual degrees of freedom from one layer to the next.}
\end{figure}

Choosing the two subsystems $A$ and $B$ of the extensive partition to be alternating blocks of $N_1,N_2$ spins, respectively, we follow the general procedure outlined above and obtain the TNS structure of the partition function, shown in Fig. 3.  Interestingly, this TNS can be viewed as a partition function for the virtual degrees of freedom (depicted by and hereafter referred to as dots) located at the ends of each block. Each ``ladder diagram" obtained by contracting physical indices (blue or green boxes in Fig. 3) plays the role of time evolution operator acting on two virtual sites. The value of the ladder can be evaluated by defining the transfer matrix $T$ acting on the auxiliary indices, defined as $T_{\alpha\beta;\gamma\delta}=\sum_\sigma M^\sigma_{\alpha\gamma}\left(M^{\sigma*}\right)_{\beta\delta}$ (see Fig.1). 
The two basic building blocks boxed in Fig. 3 are then given by $T^{N_1},T^{N_2*}$. When we permute the vertices and view $\left[T^{N_1}\right]_{\alpha\beta;\gamma\delta}$ as a mapping from auxiliary indices $\alpha\gamma$ to $\beta\delta$, it describes the ``imaginary time evolution" of the virtual sites $i,i+1$ for odd $i$ and on odd numbered rows. Similarly, $T^{N_2}$ acts on sites $j,j+1$ for even $j$ and even numbered rows.


Further simplification can be made by considering large $N_1,N_2$; in this limit, the transfer matrix $T^{N_{1(2)}}$ will be dominated by leading eigenvalues of $T$. As is well-known for MPS, by transformations in the virtual indices one can always transform $T$ into a canonical form. As we show in the Supplementary Material \cite{sm}, for the canonical $T$ in the limit of large $N_1,N_2$, neighboring dots become almost decoupled and $T^{N}$ has the form of
\beq
T^N\approx\lambda (1\otimes 1)+\xi_2^{N} (U^{L}_2 \otimes U^R_2)
\eeq
where $\lambda>0$ is a constant, $1$ is the identity operator acting on a single dot, $\xi_2$ is the second largest eigenvalue of the double tensor of the MPS, and $U_2^{L,R}$ is the corresponding left and right eigenvectors, respectively (it is nonetheless an operator acting on a single dot).  The tensor product above is that between two adjacent dots.  The key point is that MPS which represent gapped ground states have a nondegenerate largest eigenvalue of the double tensor.  This allows us to analyze the large $N$, `weak coupling' limit.

Up to a constant $\lambda$, each transfer matrix is nearly the identity. Therefore we can write $T^N\simeq \lambda \exp\left[\xi_2^N\lambda^{-1}U_2^{L}\otimes U_2^R\right]$.  The Suzuki-Trotter expansion ($e^{A}e^{B} \approx e^{A+B}$ for small $A,B$) allows us to ignore the commutator arising from the overlap of $T^{N_1}$ and $T^{N_2}$ (i.e. the green and blue boxes in Fig. 3) and write
\beq
Z\approx tr(e^{-n \tilde{H}})
\eeq
\beq
\tilde{H} \equiv &-&\sum_{\text{odd i}} \xi_2^{N_1} \lambda^{-1}(U_2^{L})_i \otimes (U_2^R)_{i+1} \\
&-&\sum_{\text{even j}} \xi_2^{N_2} \lambda^{-1}(U_2^{L})_j \otimes (U_2^R)_{j+1}, \nonumber
\eeq
up to a constant.

We now demonstrate this procedure explicitly for the BES of the Haldane phase of the spin-1 chain \cite{haldanephase}.  This is a topologically nontrivial phase protected by either time reversal, a dihedral subgroup of rotations (detailed later), or inversion symmetry \cite{pollman,terg}.  We begin by analyzing a representative of the Haldane phase: the AKLT matrix product state \cite{aklt}
\beq
M^+ = \sqrt{\frac{2}{3}} \sigma^+,
M^0 = -\frac{1}{\sqrt{3}} \sigma^z,
M^- = -\sqrt{\frac{2}{3}} \sigma^- \label{AKLT}
\eeq
Here $\pm,0$ stand for $S^z=\pm 1,0$, and $\sigma$ are the Pauli spin matrices with $\sigma^{\pm}=\frac{1}{2}(\sigma^x\pm i\sigma^y)$.  The transfer matrix is
$T_{\alpha\beta,\gamma\delta}=\frac13\sum_{i=x,y,z}\sigma^i_{\alpha\gamma}\sigma^{i*}_{\beta\delta}=\frac23\left(\delta_{\alpha\beta}\delta_{\gamma\delta}-\frac12\delta_{\alpha\gamma}\delta_{\beta\delta}\right)$. Viewing $T$ as a two-site operator acting on the auxiliary indices, we can write
\begin{eqnarray}
T=\frac12\left(1\otimes 1-\frac13\vec{\sigma}\cdot\vec{\sigma}^*\right)
\end{eqnarray}
This transfer matrix corresponds to $\xi_2=-\frac13$,~$U_2^L=\sigma^i$,~$U_2^R=\sigma^{i*}$ in the generic formula, with a three-fold degeneracy in the eigenstates. After a unitary transformation on the odd sites $\vec{\sigma}^*=-\sigma_y\vec{\sigma}\sigma_y$, the effective entanglement Hamiltonian has the Heisenberg form:
\beq
\tilde{H} \equiv -\left(-\frac{1}{3}\right)^{N_1} \sum_{\text{odd } i} P_{i,i+1}-\left(-\frac{1}{3}\right)^{N_2} \sum_{\text{even } j} P_{j,j+1}, \nonumber
\eeq
up to a constant.  Here $P$ is the projection operator of two spin-1/2s onto the singlet state.

Hence, we find that for large even $N_1,N_2$, the entanglement Hamiltonian has the same spectrum as the antiferromagnetic spin-1/2 Heisenberg chain with alternating nearest neighbor couplings.  A quantum phase transition occurs at the symmetric partition $N_1=N_2$,
where the BES describes a spin $1/2$ translation invariant Heisenberg chain. 
The quantum phase transition can also be understood in term of the original spin $1$ model. In the limit $N_1\ll N_2$, region $A$ consists of isolated blocks, while in the opposite limit $N_1\gg N_2$ it is in the Haldane phase. Therefore the phase transition is one between the Haldane phase and a trivial product state, driven by translation symmetry breaking. Indeed, this transition is known to be described by $SU(2)$ level $1$ Wess-Zumino-Witten (WZW) theory\cite{affleck,wz,w}, the same conformal field theory that describes the Heisenberg spin $1/2$ chain.

Though the above derivation applies generally to any extensive partition of any MPS representation of a gapped ground state, it is valid only in the large $N_1,N_2$ limit- an approximation which may not hold in general in higher dimensions.  Hence, we now provide a complementary analysis that 1) holds for arbitrary $N_1,N_2$ and 2) generalizes readily to higher dimensions: we view the partition function as that of a (two-dimensional) classical model.  Because we are interested in the topological phase transition expected from BES, we focus on the symmetric partition $N_1=N_2=N$ for the AKLT state.  Upon redrawing the tensor network so that each set of $N$ contracted physical indices serves as an interaction vertex between four spin-$1/2$ nodes (Fig. 4), we find that the classical partition function is that of a particular six-vertex model on a square lattice with vertex weights
\beq
V^{ij}_{kl} = \delta^i_k\delta^j_l + \lambda \delta^i_j\delta^k_l, \lambda = \frac{(-3)^N-1}{2} \nonumber
\eeq
In this model, there are two possible states on each link of the square lattice, and the weight of each configuration in the partition function is given by the product of the above vertex terms.  It is remarkable that the AKLT wavefunction ``contains'' such six-vertex models, which are exposed by BES. For the above parameters, the model is critical, equivalent to the 4-state Potts model, and described in the continuum limit by the level-1 SU(2) WZW theory \cite{Affleck}.

\begin{figure}
\includegraphics[width=3in]{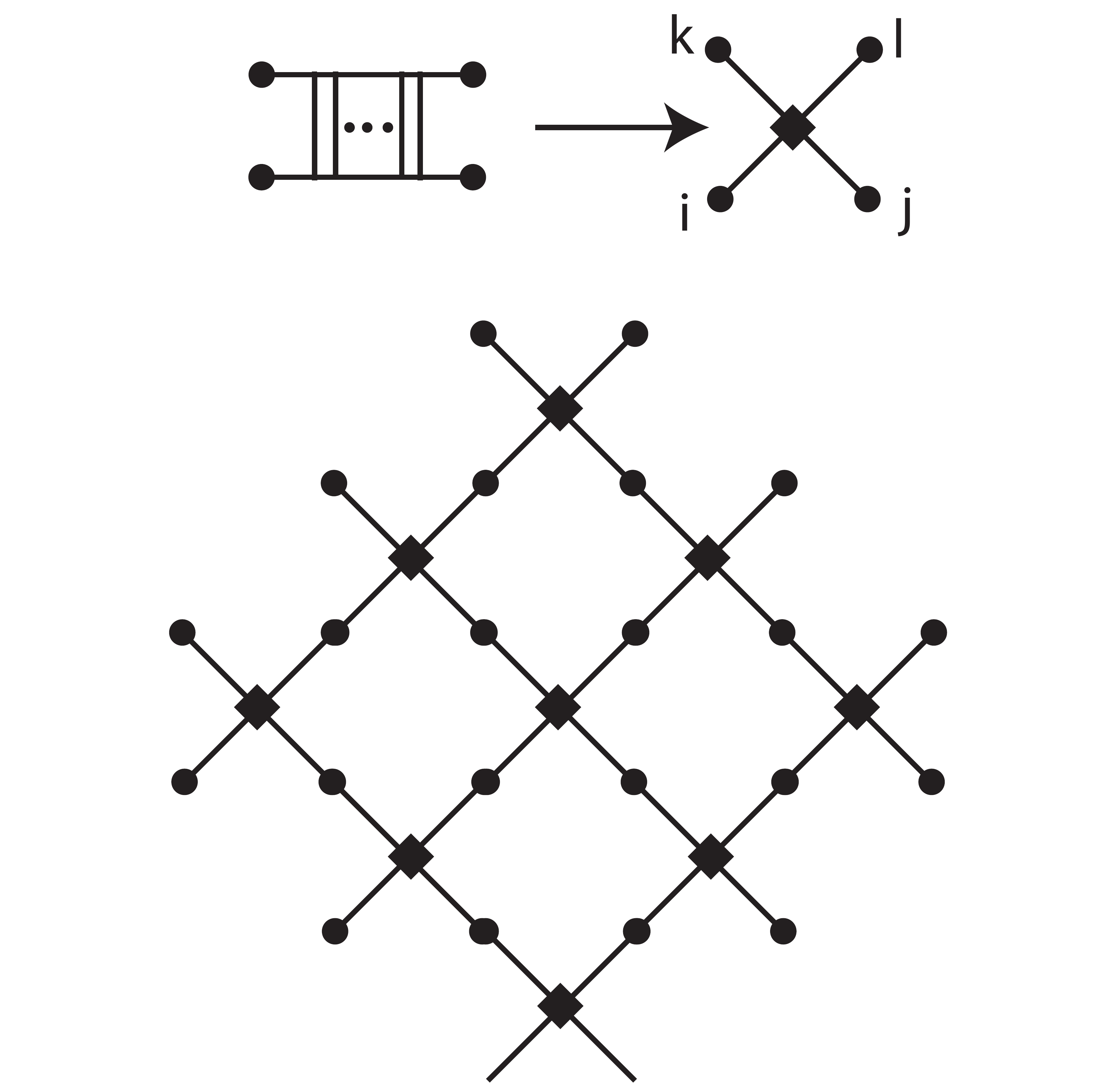}
\caption{The partition function from the entanglement Hamiltonian at the symmetric partition is identified as a classical partition function.  (top) The transfer matrix is interpreted as a vertex interaction between Ising variables (large dots).  (bottom) The resulting classical model is defined on a new lattice.}
\end{figure}

%
%
%

This particular universality class is a consequence of the $SO(3)$ symmetry of the AKLT MPS.  However, recall that the full $SO(3)$ symmetry group is not necessary to protect the topological Haldane phase.  Hence, we now consider MPS ground states which are slightly perturbed away from the AKLT state and we analyze the nature of the bulk entanglement Hamiltonian in such cases.  For this purpose, it is useful to make the spin symmetries manifest by parameterizing these MPS as \cite{liu}
\beq
M^x = a \sigma^x,
M^y = b \sigma^y,
M^z = c \sigma^z, \label{spin1}
\eeq
where $a,b,c$ are real numbers.  Compared to (\ref{AKLT}), we are simply using a new basis: $|\pm\rangle = \frac{1}{\sqrt{2}}(x\pm iy), |0\rangle = z$.  When $a=b=c$, the MPS is the $SO(3)$ symmetric AKLT state up to an overall normalization, and when two of the coefficients are equal, the MPS has at least a $U(1)$ symmetry (in the plane corresponding to those two coefficients).  Finally, when all three parameters are different, the MPS still has dihedral symmetry generated by rotations by $\pi$ about the $x,y,z$ axes.

Parameterizing small perturbations to the AKLT state by $a=1, b=1+\delta, c=1+\epsilon$, we proceed as above and find that the entanglement Hamiltonian from the symmetric partition in the large $N$ and small $\delta, \epsilon$ limit is given by the $XYZ$ spin-$1/2$ chain (see Supplementary Material \cite{sm}).  More specifically, the spin symmetry of the MPS ground state is in one to one correspondence with the symmetry of the entanglement Hamiltonian: the AKLT MPS yields the Heisenberg chain, the MPS with $U(1)$ symmetry yields the $XXZ$ chain, and the most general MPS of the type (\ref{spin1}) yields the $XYZ$ chain.  Since the $XYZ$ chain is either critical or spontaneously orders along a direction \cite{XYZ1,XYZ2,XYZ3}, the corresponding entanglement Hamiltonian is either critical or has ground state degeneracy, both of which are consistent with our general arguments \cite{BES}.

Because of the two free parameters in the MPS, the resulting entanglement Hamiltonian will generically be in the gapped part (with ground state degeneracy) of the $XYZ$ phase diagram.  If the MPS is fine tuned so that the corresponding entanglement Hamiltonian lies on one of the critical lines, then the universality class can be simply extracted from the $XYZ$ model.  All critical lines in the $XYZ$ phase diagram map to the $XXZ$ chain \cite{XYZ3}, which in turn can be mapped via bosonization to (critical) Luttinger liquids with Luttinger parameter depending on the $XXZ$ anisotropy.  We conclude that such critical theories describe the transition of the Haldane phase to a topologically trivial dimerized phase.

Alternatively, for finite $N$, we can also map the partition function of the entanglement Hamiltonian for this wider class of MPS to classical models in two dimensions, as we derived a six-vertex model from the AKLT state.  We find that the general class of MPS of the form (\ref{spin1}) maps into eight-vertex models \cite{XYZ1}, though in some cases the weights of some configurations may be negative.

By using tensor networks to construct the partition function of the critical bulk entanglement Hamiltonian, we gain much insight into the topological phase transition revealed by BES.  Previously, topological phase transitions have been described by the condensation of fractionalized excitations or the delocalization of edge states \cite{criticaledge}; in both cases, the original degrees of freedom of the model are overshadowed by emergent ones.  Thanks to the partition function and the tensor network framework, we have a clear picture of the virtual degrees of freedom interacting and giving rise to the phase transition.  As a byproduct of this procedure, we attain classical lattice models of quantum critical points; as an example, we unearthed critical six-vertex models from the AKLT wavefunction.  We note that this tensor network implementation of BES generalizes readily to higher dimensions, in which interesting quantum-classical mappings may await.

{\it Acknowledgement:}
We thank T. Senthil for informing us of the relation between the Heisenberg model and the 4-state Potts model.  TH is supported by NSF Graduate Research Fellowship No. 0645960. LF is supported by the DOE Office of Basic Energy
Sciences, Division of Materials Sciences and Engineering
under award DE-SC0010526. XLQ is supported by the National Science Foundation through the grant No. DMR-1151786.

{\it Note added:}
After completion of this work (see \cite{liang} for Simons Center presentation or \cite{tim} for Simons Institute poster), we noticed numerical results \cite{related} for a fixed extensive partition of the AKLT chain in which alternating blocks of two sites are traced out.  These results support our conclusions.

\section{Supplementary Material}

\subsection{Derivation of Effective Entanglement Hamiltonian}

Given an MPS representing a unique gapped ground state wavefunction in one dimension, we derive the effective entanglement Hamiltonian from an extensive partition consisting of alternating blocks of $N_1,N_2$ sites in the limit of large $N_1$, $N_2$.  The first step is evaluating the transfer matrices $T^{N_1},T^{N_2}$.  To proceed, first consider the double tensor $T$, given by a single unit of the transfer matrix shown in the box of Fig. 1 of main text.  Diagonalizing,
\beq
T_{\alpha\beta;\gamma\delta}=U^L_{\alpha\beta,n} \xi_n U^R_{\gamma\delta,n}
\eeq
where $U$ is a unitary matrix, and $n$ indexes the eigenvalues from highest $\xi_1$ to lowest.  Because the MPS is assumed to be a ground state with a gap above it, the largest eigenvalue of $E$ is non degenerate, and for simplicity we will assume the second largest eigenvalue is also non degenerate (the derivation below can be easily generalized).  


The double tensor allows a basis transformation such that the largest eigenvalue is 1 and the corresponding eigenvectors have the form 
\beq
U^R_{1,\gamma\delta}&=&\delta_{\gamma\delta} \\
U^L_{\alpha\beta,1}&=&\delta_{\alpha\beta} \lambda_{\alpha}
\eeq
where $\lambda_{\alpha}>0$ \cite{RG}.
Exponentiating the double tensor along the horizontal direction and keeping nontrivial leading order gives
\beq
T^N&=&\delta_{\alpha\beta}\delta_{\gamma\delta}\lambda_{\alpha}+\xi_2^{N} U^L_{\alpha\beta,2} U^R_{2,\gamma\delta}\\
&\approx&e^{-\tilde{H}} \\
\tilde{H}&=&-\log{\lambda_{\alpha}} \otimes 1-\xi_2^N \lambda^{-1}_{\alpha}U_2^L \otimes U^R_2.
\eeq
The Suzuki-Trotter expansion ($e^{A}e^{B} \approx e^{A+B}$ for small $A,B$) allows us to write the total partition function as
\beq
Z\approx tr(e^{-\beta \tilde{H}}) 
\eeq
\beq
\tilde{H} \equiv &-&\sum_{odd i} \xi_2^{N_1} \lambda^{-1}_{\alpha}(U_2^L)_i \otimes (U^R_2)_{i+1} \\
&-&\sum_{even j} \xi_2^{N_2} \lambda^{-1}_{\alpha}(U_2^L)_j \otimes (U^R_2)_{j+1}, \nonumber
\eeq
up to a constant.  (Since each block of the partition function is close to the identity, the lack of commutation between blocks from different layers is higher order in our expansion).

\subsection{Effective Entanglement Hamiltonian for AKLT}
Here we apply the above general derivation to the AKLT MPS; the techniques used will be applied in the next section to generalized AKLT states.  A straightforward calculation for the AKLT state shows that $T^N$, interpreted as acting on two indices of one row to yield a state in the next row, is given by 
\beq
T^{N} = \frac{1}{2}\left(1-\left(-\frac{1}{3}\right)^{N}\right)I + 2\left(-\frac{1}{3}\right)^{N} P,
\eeq
where $I$ is the identity operator and
\beq
P\equiv \frac{1}{2} |\uparrow \uparrow + \downarrow \downarrow\rangle\langle \uparrow \uparrow + \downarrow \downarrow|.
\eeq 

Performing a change of basis $|\uparrow\rangle \rightarrow |\downarrow\rangle, |\downarrow\rangle \rightarrow -|\uparrow\rangle$ on the second spin-1/2 of each pair, we see that $P$ is simply the projector onto the singlet.  Thus, the spin-1/2s have (anti)ferromagnetic interaction for (even) odd $N$.  

In the large-$N$ limit, $T^N\approx (1/2)(I+ 4(-\frac{1}{3})^N P) \approx (1/2) e^{4(-\frac{1}{3})^N P}$.  In this limit, the Suzuki-Trotter expansion ($e^{A}e^{B} \approx e^{A+B}$ for small $A,B$) allows us to write
\beq
Z\approx tr(e^{-\beta \tilde{H}}) 
\eeq
\beq
\tilde{H} \equiv -\left(-\frac{1}{3}\right)^{N_1} \sum_{odd i} P_{i,i+1}-\left(-\frac{1}{3}\right)^{N_2} \sum_{even j} P_{j,j+1}, \nonumber
\eeq
up to a constant.

\subsection{BES of Perturbed AKLT}

Here we consider the BES of a spin 1 chain in the Haldane phase with a smaller symmetry group than that of the AKLT MPS.  It is convenient to use the following parameterization \cite{akltgeneral} for each MPS building block: $\sigma_x, b\sigma_y, c\sigma_z$.   

If $b=c=1$, we recover the AKLT MPS, with $SO(3)$ symmetry.  If $b=1$ (or $c=1$), there is $U(1)$ symmetry.  If all three are different, there is dihedral $Z_2 \times Z_2$ symmetry.  For example, under $\pi$ rotation about the $x$-axis, $\sigma_x \rightarrow \sigma_x, \sigma_{y,z} \rightarrow -\sigma_{y,z}$.

The double tensor formed by contracting physical indices is
\beq
T=\sigma_x \otimes \sigma_x + b^2 \sigma_y \otimes \sigma^*_y + c^2 \sigma_z \otimes \sigma_z.
\eeq

%
%
%

Consider small perturbations around the AKLT MPS: $b=1+\delta/2, c=1+\epsilon/2$.  

Then to leading order
$$T=\frac{1}{3}
\left( \begin{array}{cccc}
1+\epsilon & 0 & 0 & 2+\delta \\
0 & -1-\epsilon & -\delta & 0 \\
0 & -\delta & -1-\epsilon & 0 \\
2+\delta & 0 & 0 & 1+\epsilon 
\end{array} \right)
$$

and the full transfer matrix for a block of $N$ sites is

\beq
T^N&=&\left( \begin{array}{cccc}
f & 0 & 0 & j \\
0 & g & k & 0 \\
0 & k & g & 0 \\
j & 0 & 0 & f \label{transfer}
\end{array} 
\right)
\eeq

where

\beq
f(N,\delta,\epsilon)&=&\frac{1}{2} \frac{1}{3^N}\left((-1+\epsilon-\delta)^N + (3+\epsilon+\delta)^N\right) \nonumber \\
g(N,\delta,\epsilon)&=&\frac{1}{2} \frac{1}{3^N}\left(-(-1+\epsilon-\delta)^N + (3+\epsilon+\delta)^N\right) \nonumber \\
j(N,\delta,\epsilon)&=& \frac{1}{2} \frac{1}{3^N}\left((-1-\epsilon+\delta)^N + (-1-\epsilon-\delta)^N\right) \nonumber \\
k(N,\delta,\epsilon)&=& \frac{1}{2} \frac{1}{3^N}\left(-(-1-\epsilon+\delta)^N + (-1-\epsilon-\delta)^N\right) \nonumber
\eeq

As done for the AKLT state in the above section, we now perform a change of basis $|\uparrow\rangle \rightarrow |\downarrow\rangle, |\downarrow\rangle \rightarrow -|\uparrow\rangle$ on the second spin-1/2 of each pair.

%
%

After this unitary transformation,
%
\beq
\tilde{T}^N&=&\left( \begin{array}{cccc}
g & 0 & 0 & -k \\
0 & f & -j & 0 \\
0 & -j & f & 0 \\
-k & 0 & 0 & g
\end{array} 
\right) \\
&=&\frac{f+g}{2}I-\frac{j+k}{2} \sigma_x \otimes \sigma_x \\
&-& \frac{j-k}{2}\sigma_y \otimes \sigma_y -\frac{f-g}{2} \sigma_z \otimes \sigma_z \nonumber
\eeq
%

To leading order in $1/N$, 

\beq
\tilde{T}^N=\frac{1}{2}\frac{1}{3^N} &&((3+\epsilon+\delta)^N I \\
&&-(-1-\epsilon-\delta)^N \sigma_x \otimes \sigma_x \nonumber \\
&&-(-1-\epsilon+\delta)^N \sigma_y \otimes \sigma_y \nonumber \\
&&-(-1+\epsilon-\delta)^N \sigma_z \otimes \sigma_z) \nonumber
\eeq


We see that the transfer matrix is primarily the identity in the large N limit, with corrections in N given by Heisenberg exchange and corrections in the perturbations away from AKLT given by anisotropic terms.

Hence the effective entanglement Hamiltonian given by the transfer matrix acting on one pair of sites and then the adjacent pair on the next layer of the partition function is 

\beq
\tilde{H} = \sum_{i,j} \frac{1}{(3+\epsilon+\delta)^N} &&((-1-\epsilon-\delta)^N \sigma^i_x \otimes \sigma^j_x \nonumber \\
&&(-1-\epsilon+\delta)^N \sigma^i_y \otimes \sigma^j_y  \nonumber\\
&&(-1+\epsilon-\delta)^N \sigma^i_z \otimes \sigma^j_z) \nonumber \label{EH}
\eeq 
up to a constant. 

As explained in the text, the partition function of the entanglement Hamiltonian from an extensive partition of the AKLT state maps to the six-vertex model with a special set of couplings which gives $SU(2)$ level 1 WZW in the continuum.  In general, when the spin symmetry is reduced to the dihedral group, all entries of the transfer matrix (\ref{transfer}) are nonzero and the corresponding classical model is an eight-vertex model.


\begin{thebibliography}{10}
\bibitem{irreducible}
A topological state is dubbed irreducible if it cannot be written as the direct product of two states with identical, nontrivial topological invariants or order.  A (counter)example is a state with Chern number $\nu=1(2)$.
\end{thebibliography}
\end{document}